\documentclass{elsart}
\usepackage{graphicx}
\usepackage{amssymb}
\usepackage{epsfig}
\def\Journal#1#2#3#4{{#1} {#2} (#3) #4}
\def\NPA{Nucl. Phys. A}
\def\NPB{Nucl. Phys. B}
\def\PLB{Phys. Lett.  B}
\def\PRL{Phys. Rev. Lett.}
\def\PRC{Phys. Rev. C}
\def\PRD{Phys. Rev. D}

\def\ZPC{Z. Phys. C}

\def\EPJC{Eur. Phys. J. C}
\def\JPG{J. Phys. G}

\newcommand{\ud}{\mathrm{d}}

\newcommand{\AmS}{{\protect\the\textfont2
  A\kern-.1667em\lower.5ex\hbox{M}\kern-.125emS}}

\begin{document}

\begin{frontmatter}
% declarations for front matter

\title{Statistical hadronization of charm in heavy-ion collisions
at SPS, RHIC and LHC}

\author[gsi]{A.~Andronic}, %\thanksref{info}},
\author[gsi]{P.~Braun-Munzinger},
\author[wro,bie]{K.~Redlich},
\author[hei]{J.~Stachel}

\address[gsi]{Gesellschaft f\"ur Schwerionenforschung, GSI, \\D-64291
Darmstadt, Germany}
\address[wro]{Fakult\"at f\"ur Physik, Universit\"at Bielefeld,\\
Postfach 100 131, D-33501 Bielefeld, Germany}
\address[bie]{ Institute of Theoretical Physics, University of Wroc\l aw,\\
PL-50204 Wroc\l aw, Poland}
\address[hei]{ Physikalisches Institut der Universit\"at Heidelberg,\\
D-69120 Heidelberg, Germany}
\begin{abstract}
We study the production of charmonia and charmed hadrons in
nucleus-nucleus collisions at SPS, RHIC, and LHC energies within the
framework of the statistical hadronization model.  Results from this
model are compared to the measured yields and centrality dependence of
$J/\psi$ production at SPS and RHIC energies.  We explore and contrast
the centrality dependence of the production of mesons with open and
hidden charm at SPS, RHIC and LHC.  The sensitivity of the results to
various input parameters is analyzed in detail for RHIC energy.
\end{abstract}

\end{frontmatter}

\section{Introduction}

 The concept of a possible thermal origin of charmonia was initially
 introduced \cite{marek} to explain the measurement of the
 J/$\psi$/hadron ratio in nuclear collisions. Because of their large
 mass and small production cross section at thermal energies charm
 quarks are, however, not likely to be thermally produced. On the
 other hand, significant
 production of charm quark pairs takes place in initial hard
 collisions. This led to the idea of statistical hadronization of charm quarks
\cite{pbm1,pbm2}, which has sparked an intense activity in this field
\cite{gor1,gra1,gor2,kos,gra2}. At about the same time an
independent effort \cite{the1,the2} based on a kinetic model has
been developed to study charm production in heavy ion collisions at
 collider energies.
Initial interest focussed on the available SPS data on $J/\psi$
production in Pb--Pb collisions.  As we stress  below, these data
can indeed be described  within the statistical approach, but only
when assuming the  charm production cross section to be  enhanced beyond the
perturbative QCD (pQCD) predictions.

 The largest differences between the results of the statistical
coalescence scenario and earlier  models are expected at collider energies. 
For example, for central Au-Au collision at RHIC energy,  the
Satz--Matsui approach \cite{satz} predicts a  very strong (up to
a factor of 20 \cite{vogt99}) suppression of  $J/\psi$  yields as
compared to a direct production.  In the present approach this
suppression is overcome by statistical recombination of $J/\psi$
mesons from the initially produced $c \bar c$ pairs, so that much larger
yields are expected.\footnote{ The absence of $J/\psi$
suppression at RHIC energy is also predicted by the kinetic model
 \cite{the1,the2}.}
We therefore focus in this letter on the predictions of the model
for open and hidden charm meson yields at RHIC (Au--Au) and LHC
(Pb--Pb) energies, with  particular emphasis  on the centrality
dependence of rapidity densities at RHIC. We, furthermore, present
predictions for the energy dependence of the production of hadrons
with open and hidden charm.

\section{Model description and input parameters}

The model assumes that all charm quarks are produced in primary
hard collisions and equilibrate\footnote{This implies thermal, but
not chemical equilibrium for charm quarks.} in the quark-gluon
plasma (QGP). In particular, no $J/\psi$ mesons are preformed in
the QGP, implying  complete color screening \cite{satz}. There are
indications from recent lattice QCD calculations 
that, already at SPS energies, J/$\psi$ mesons can be dissociated in
a deconfined medium via collisions with thermal gluons
\cite{kar,wong1}. As noted earlier \cite{pbm1}, the cross sections 
of charmed hadrons are too small to allow for their chemical equilibration 
in a hadronic gas.
We assume that open and hidden charm hadrons are formed at chemical 
freeze-out according to statistical laws. Consistent with the fact that,
at SPS energy and beyond, chemical freeze-out appears to be at the phase 
boundary \cite{pbm3}, our model implies that a QGP phase was a stage 
in the evolution of the fireball. The analysis of $J/\psi$ spectra at 
SPS \cite{bug} lends further support to the statistical hadronization 
picture where $J/\psi$ decouples at chemical freeze-out. 
A recent analysis of electron spectra at RHIC \cite{bat} also
strenghtens the case for an early thermalization of heavy quarks.
However, in that analysis it was pointed out that both the hydrodynamical 
approach and PYTHIA reproduce the measured single-electron spectra.
% is for the moment a puzzling observation. 
As suggested meanwhile \cite{nag}, the measurement of the elliptic flow 
($v_2$) of open charm hadrons and $J/\psi$ may help to disentangle 
the two scenarios. Recent work \cite{limo}
in extracting the quark $v_2$ from the one of hadrons may help towards 
this goal.
Most important though, in our view, is a high-precision direct measurement
of open charm. As expected, and as seen in ref. \cite{bat}, 
the $p_t$ spectra predicted by PYTHIA and hydro are different in detail at 
low $p_t$ and differ manifestly at high $p_t$ ($p_t \gg$ mass of charm quark).
On the other hand, a recent theoretical analysis \cite{dok} indicates 
that charm quarks might not thermalize quickly because of their large mass. 

In  statistical models  charm production needs generally to be treated
within the framework of canonical thermodynamics \cite{cle}. 
Thus, the charm balance equation  required \cite{pbm1,pbm2} during
hadronization is expressed as:
\begin{equation}
N_{c\bar{c}}^{dir}=\frac{1}{2}g_c N_{oc}^{th}
\frac{I_1(g_cN_{oc}^{th})}{I_0(g_cN_{oc}^{th})} + g_c^2N_{c\bar c}^{th}.
\label{aa:eq1}
\end{equation}
Here $N_{c\bar{c}}^{dir}$  is the number of directly produced
$c\bar{c}$ pairs and  $I_n$ are  modified Bessel functions. In
the fireball of volume $V$  the total number of open
$N_{oc}^{th}=n_{oc}^{th}V$ and hidden  $N_{c\bar c}^{th}=n_{c\bar c}^{th}V$  
charm hadrons  are computed from their grand-canonical densities 
$n_{oc}^{th}$ and $n_{c\bar c}^{th}$, respectively. 
The densities of different particle species in the grand canonical ensemble 
are calculated following the statistical model introduced in \cite{heppe}.
All known charmed mesons and hyperons and their decays \cite{pdg} 
have to be included in the calculations.

The balance equation (\ref{aa:eq1})
defines a fugacity parameter $g_c$ that accounts for deviations of charm 
multiplicity  from the value that is expected in complete chemical equilibrium.
The yield  of open charm mesons and hyperons $i$ and of charmonia $j$ is 
obtained from:
\begin{equation}
N_i=g_cN_i^{th}\frac{{I_1(g_c N_{oc}^{th})}}{{I_0(g_c N_{oc}^{th})}}
\quad \mathrm{and} \quad N_j=g_c^2 N_j^{th}. \label{aa:eq2}
\end{equation}
The above model for charm production and hadronization can be only
used  \cite{pbm1,sor}  if  the number of
participating nucleons $N_{part}$  is sufficiently large. Taking 
into account the measured dependence of the relative yield of
$\psi'$ to $J/\psi$ on centrality \cite{na38} in Pb--Pb collisions
at SPS energy the model appears appropriate  \cite{pbm1,sor} if
$N_{part}>$100.

To calculate the yields of open and hidden charm hadrons for a
given centrality and collision energy  one needs to fix a set of
parameters in Eq.(1) and (2):

i) A constant temperature of 170 MeV and a baryonic chemical
potential $\mu_b$ according to the parametrization $\mu_b=1270[{\rm
MeV}]/(1+\sqrt{s_{NN}}[{\rm GeV}]/4.3)$ \cite{pbm4}
%($\sqrt{s_{NN}}$ is in GeV)
are used for our calculations. For SPS and RHIC energies these
thermal parameters are consistent, within statistical errors, with
those required to describe experimental data on  different hadron
yields \cite{heppe,pbmn}.
In Table~\ref{aa:tab0} we present the values of these input parameters
and the resulting chemical potentials for isospin, strangeness and 
charmness for the canonical ensemble calculations at SPS, RHIC and LHC.

\begin{figure}[htb]
%exec plot#s opt=ch
\centering\includegraphics[width=.65\textwidth]{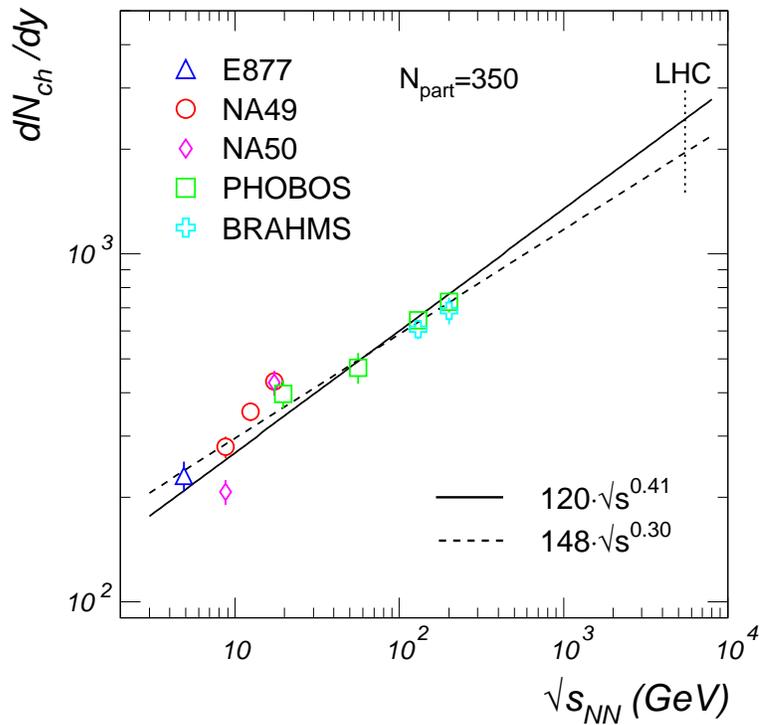}
\caption{Experimental values of charged particle yields at
midrapidity as a function of collisions energy for central
collisions.  The data points are from experiments at AGS (E877
\cite{e877}), SPS (NA49 \cite{na49}, NA50 \cite{na50a}) and RHIC
(PHOBOS \cite{pho}, BRAHMS \cite{bra}).  The continuous and dashed
lines are power law dependences (see text), the dotted line marks
the full LHC energy for Pb--Pb collisions ($\sqrt{s_{NN}}$=5.5
TeV).} \label{aa:fig0}
\end{figure}

\begin{table}[htb]
\caption{Input values of $T$ and $\mu_b$ and output chemical potentials
for isospin, strangeness and charmness for canonical ensemble calculations 
at SPS, RHIC and LHC.}
\label{aa:tab0}
\begin{tabular}{c|ccc} %\hline
$\sqrt{s_{NN}}$ (GeV) &  17.3  &  200 &  5500  \\ \hline
$T$ (MeV)        &  170  &  170  & 170  \\
$\mu_b$ (MeV)    &  253  &  27   & 1  \\ \hline
$\mu_{I_3}$ (MeV)& -10.1 & -0.96 & -0.037 \\
$\mu_S$ (MeV)    & -68.6 &  6.70 &  0.304 \\
$\mu_C$ (MeV)    & -41.8 & -3.92 & -0.156 \\

\end{tabular}
\end{table}

ii) The volume of the fireball corresponding to a slice
of one rapidity unit at midrapidity, $V_{\Delta y=1}$, is obtained from 
the charged particle rapidity density $\ud N_{ch}/\ud y$, via the relation 
$\ud N_{ch}/\ud y=n_{ch}^{th}V_{\Delta y=1}$, where $n_{ch}^{th}$ is the 
charged particle density computed within the thermal model. 
The charged particle rapidity densities (and total yields in case of SPS) 
are taken from experiments at SPS \cite{na49} and RHIC \cite{pho}
and extrapolated to LHC energy. 
In Fig.~\ref{aa:fig0} we show a compilation of experimental data on
$\ud N_{ch}/\ud y$ at midrapidity for central collisions in the energy 
range from AGS \cite{e877} up to RHIC\footnote{A constant Jacobian of 
1.1 has been used to convert the $\ud N_{ch}/\ud\eta$ data to 
$\ud N_{ch}/\ud y$.}.
The continuous line  in Fig.~\ref{aa:fig0} is the prediction
\cite{esk1} of the saturation model, the dashed line is a flatter
power law dependence, both arbitrarily normalized. We use the
$(\sqrt{s_{NN}})^{0.3}$ dependence of  $\ud N_{ch}/\ud y$ to get
the value of $\ud N_{ch}/\ud y\simeq$2000 for LHC energy.
Throughout this paper we define central collisions by the average value 
$N_{part}$=350, which roughly corresponds to the topmost 5\% of the 
nuclear cross section.
For the centrality dependences we assume that the volume of the fireball
is proportional to $N_{part}$.

iii) The yield of open charm $N_{c\bar{c}}^{dir}$ is taken from
next-to-leading order (NLO) pQCD calculations for p--p collisions
\cite{vogt1} and scaled to nucleus--nucleus collision via the
nuclear overlap function \cite{esk,dar} $T_{AA}$.  For a given
centrality:
$N_{c\bar{c}}^{dir}(N_{part})= \sigma({pp}\rightarrow {c\bar{c}})
T_{AA}(N_{part})$.
The pQCD calculations with the  MRST HO parton distribution
function (PDF) are used here \cite{vogt1}. 
We note that the pQCD results, in particular at LHC energy, are sensitive 
on the choice of PDF and/or charm quark mass value \cite{vogt1}.

The input values $\ud N_{ch}/\ud y$ and $\ud N_{c\bar{c}}^{dir}/\ud y$ 
and the corresponding volume at midrapidity and enhancement factor are 
summarized in Table~\ref{aa:tab1} for model calculations for different 
collision energies.

\begin{table}[htb]
\caption{Input ($\ud N_{ch}/\ud y$ and $\ud N_{c\bar{c}}^{dir}/\ud y$) 
and output ($V_{\Delta y=1}$ and $g_c$) parameters for model calculations 
at top SPS, RHIC and LHC for central collisions.}
\label{aa:tab1}
\begin{tabular}{c|ccc} %\hline
$\sqrt{s_{NN}}$ (GeV) &  17.3  &  200 &  5500  \\ \hline
$\ud N_{ch}/\ud y$  &  430  &  730  &  2000  \\ 
$\ud N_{c\bar{c}}^{dir}/\ud y$  & 0.064 & 1.92  & 16.8  \\ \hline
$V_{\Delta y=1}$ (fm$^{3}$)     &  861  & 1663  & 4564 \\
$g_c$                           & 1.86  & 8.33  & 23.2 \\ 
\end{tabular}
\end{table}

\section{Model results and predictions}

\begin{figure}[htb]
%plot#jpsi-npart-sps xk=-1 ym=0.3
\centering\includegraphics[width=.63\textwidth]{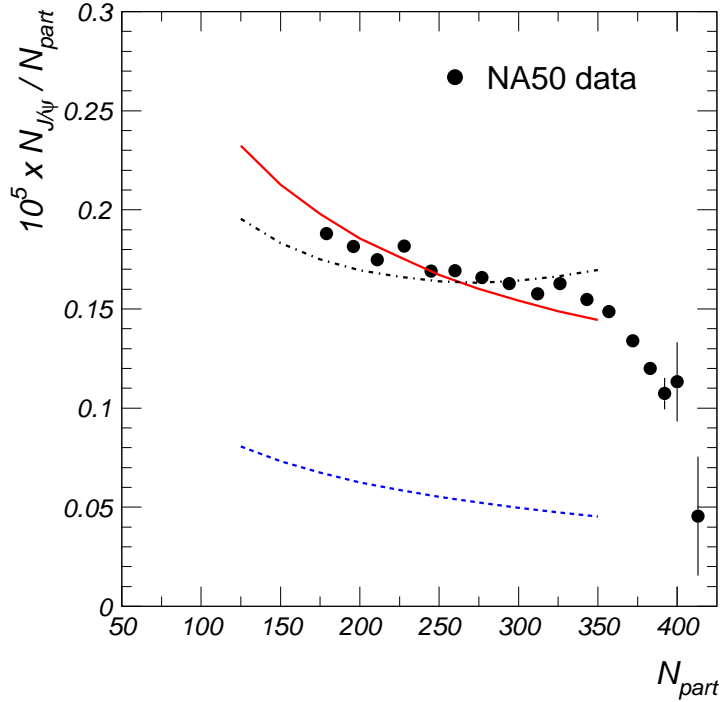}
\caption{The centrality dependence of $J/\psi$ production at SPS.
Model predictions are compared to 4$\pi$-integrated NA50 data \cite{na50,gos}.
Two curves for the model correspond to the values of 
$N_{c\bar{c}}^{dir}$ from NLO calculations (dashed line) and scaled up by a 
factor of 2.8 (continuous line). 
The dash-dotted curve is obtained when considering the possible NA50 
$N_{part}$-dependent charm enhancement over their extracted p--p cross section
\cite{na50b} (see text).}
\label{aa:fig1}
\end{figure}

 We first compare predictions of the model to 4$\pi$-integrated
$J/\psi$ data at the SPS energy. The NA50 data \cite{na50,gos} are
replotted in Fig.~\ref{aa:fig1} as outlined in \cite{pbm2}. The
model results shown in Fig.~\ref{aa:fig1} are going beyond that of
\cite{pbm2} since  we have included a complete set of charmed
mesons and baryons as well as updated the values of the charm
production cross section and the volume of the fireball.
Recent results  \cite{na49} of the NA49 Collaboration  on total charged 
particle multiplicity in central collisions yield $N_{ch}=1533$, 
leading  to a fireball total volume V=3070~fm$^3$ and to a corresponding 
total yield of thermal open charm pairs of $N_{oc}^{th}$=0.98.
 This is to be contrasted with $N_{c\bar{c}}^{dir}$=0.137 from NLO 
calculations \cite{vogt1}, leading to a value of $g_c$=0.78.
Although $g_c$ is here close to unity, this obviously does not
indicate that charm production appears at chemical equilibrium, 
as the suppression factor is a strongly varying function of the
collision energy. We have already indicated that, within the time
scales available in heavy ion collisions, the chemical equilibration 
of charm  is very unlikely both  in confined and deconfined media
\cite{pbm-redlich}.

In Fig.~\ref{aa:fig1} we show the comparison between the results
of our model and NA50 data for two different values of
$N_{c\bar{c}}^{dir}$: from NLO calculations \cite{vogt1} and
scaled up by a factor of 2.8. 
%The observed centrality dependence of $J/\psi$ is well reproduced 
%using the NLO cross sections for charm production scaled by the 
%nuclear overlap function. 
Using the NLO cross sections for charm production scaled by the 
nuclear overlap function, the model understimates the measured yield.
To explain the overall magnitude of the data, one needs to 
increase the $N_{c\bar{c}}^{dir}$ yield  by a factor of 2.8 as compared 
to NLO calculations. We mention in this context that the observed
\cite{na50b} enhancement of the di-muon yield at intermediate
masses has been interpreted 
as a possible indication for an anomalous increase of the charm production 
cross section \cite{wong}.
A third calculation (resulting in the dash-dotted line in 
Fig.~\ref{aa:fig1}) is using the NLO cross section scaled-up by 1.6,
which is the ratio of the open charm cross section estimated 
by NA50 for p--p collisions at 450 GeV/c \cite{na50b} 
and the present NLO values.
For this case the $N_{part}$ scaling is not the overlap function, but
is taken according to the measured di-muon enhancement as a function
of $N_{part}$  \cite{na50b}.
The resulting $J/\psi$ yields from the statistical model are on
average in agreement with the data, albeit with a flatter centrality 
dependence than by using the nuclear overlap function. 
Thus our charm enhancement factor of 2.8 needed to explain the $J/\psi$ 
data is very similar to the factor needed to explain the intermediate mass 
dilepton enhancement assuming that it arises exclusively from charm
enhancement \cite{na50b}.
 We note, however, that other plausible explanations exist of the 
observed enhancement in terms of thermal radiation \cite{rapp,gall}.

The drop of the $J/\psi$ yield per participant observed in the
data for $N_{part}>$350 (see Fig.~\ref{aa:fig1}) is currently
understood in terms of further $J/\psi$ dissociation due to 
energy density fluctuations for a given overlap geometry \cite{bla}.  
Alternatively, the drop has been connected with a possible trigger bias 
in the data \cite{cap}.
Note that this second drop is only visible in the minimum bias
analysis \cite{na50}.
These aspects are not included in the current model, however  when
being incorporated  the  $J/\psi$  yield for very central
collisions can be as well reproduced  \cite{kos,gra2}.

\begin{table}[htb]
\caption{Mid-rapidity densities for open and hidden charm hadrons,
calculated for central collisions at SPS, RHIC and LHC.}
\label{aa:tab2}
\begin{tabular}{c|ccc}
$\sqrt{s_{NN}}$ (GeV) &  17.3  &  200 &  5500  \\ \hline
$\ud N_{\mathrm{D}^+}/\ud y$   & 0.010 & 0.404 & 3.56 \\
$\ud N_{\mathrm{D}^-}/\ud y$   & 0.016 & 0.420 & 3.53 \\
$\ud N_{\mathrm{D}^0}/\ud y$   & 0.022 & 0.888 & 7.80 \\
$\ud N_{\mathrm{\bar D}^0}/\ud y$ & 0.035 & 0.928 & 7.82 \\
$\ud N_{\mathrm{D_s}^+}/\ud y$ & 0.012 & 0.349 & 2.96 \\
$\ud N_{\mathrm{D_s}^-}/\ud y$ & 0.009 & 0.338 & 2.95 \\
$\ud N_{\Lambda_c}/\ud y$      & 0.014 & 0.153 & 1.16 \\
$\ud N_{\bar \Lambda_c}/\ud y$ & 0.0012 & 0.117 & 1.15 \\
$\ud N_{J/\psi}/\ud y$  &  2.55$\cdot$10$^{-4}$ & 0.011 & 0.226 \\
$\ud N_{\psi'}/\ud y$ & 0.95$\cdot$10$^{-5}$ & 3.97$\cdot$10$^{-4}$ & 8.46$\cdot$10$^{-3}$
\\ %\hline
\end{tabular}
\end{table}

We turn now to discuss our model predictions for charmonia and
open charm production at collider energies and compare them with
the  results obtained at SPS.  
Notice that from now on we focus on rapidity densities,
which are the relevant observables at the colliders.
In Table~\ref{aa:tab2} we summarize the yields for a selection of hadrons 
with open and hidden charm. All predicted yields increase strongly with 
beam energy, reflecting the increased charm cross section and the
concomitant importance of statistical recombination. 
Also, ratios of open charm hadrons evolve with increasing energy, 
reflecting the corresponding decrease in the charm chemical potential.

\begin{figure}[hbt]
\centering\includegraphics[width=.57\textwidth]{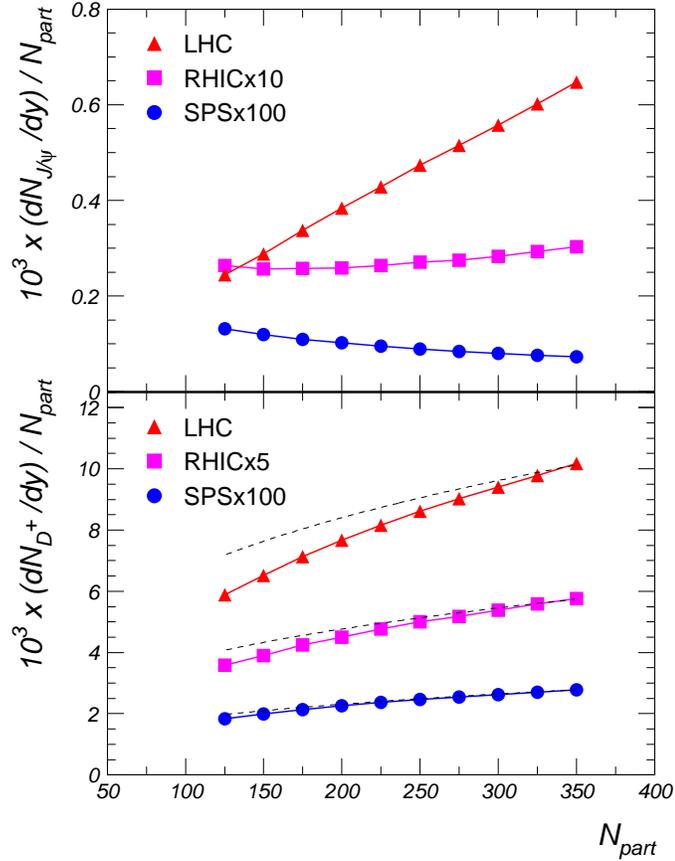}

\vspace{-.5cm}
\caption{Centrality dependence of rapidity densities of $J/\psi$
(upper panel) and D$^+$ (lower panel) mesons per participant at SPS,
RHIC and LHC.  Note the scale factors for RHIC and SPS energies.  The
dashed lines in the lower panel represent $N_{part}^{1/3}$ dependences
normalized for $N_{part}$=350.  }
\label{aa:fig2}
\end{figure}

Model predictions for the centrality dependence of $J/\psi$ and D$^+$
rapidity densities normalized to $N_{part}$ are shown in
Fig.~\ref{aa:fig2}.  The results for $J/\psi$ mesons exhibit, in
addition to the dramatic change in magnitude, a striking change in the
shape of the centrality dependence. In terms of the model this change
is a consequence of the transition from the canonical to the
grand-canonical regime \cite{pbm2}.  For  D$^+$-mesons, the expected
approximate scaling of the ratio D$^+$/$N_{part}$ 
$\propto N_{part}^{1/3}$ (dashed lines in Fig.~\ref{aa:fig2}) is only roughly
fulfilled due to departures of the nuclear overlap function from the
simple $N_{part}^{4/3}$ dependence.

The results summarized in Table~\ref{aa:tab2} and shown in
Fig.~\ref{aa:fig2} obviously depend on two input parameters,
$\ud N_{ch}/\ud y$ and $\ud N_{c\bar{c}}^{dir}/\ud y$. 
For LHC energy, neither one of these parameters is well known. 
An increase of charged particle  multiplicities by up to a factor 
of three beyond  our ``nominal'' value $\ud N_{ch}/\ud y$=2000 for
central collisions is conceivable. The  saturation model, e.g.,
leads to a prediction of \cite{esk1} $\ud N_{ch}/\ud y\simeq$2300.
However,  
due to quite large uncertainties on the  amount of shadowing at
LHC energy, these results may be still modified. The yield of 
$\ud N_{c\bar{c}}^{dir}/\ud y$ is also not well known at LHC energy. 
Although these uncertainties affect considerably the magnitude of the
predicted yields,  their centrality dependence remains
qualitatively unchanged: the yields per participant are increasing
functions of $N_{part}$.  We also note here that, while detailed
predictions differ significantly, qualitatively similar results
(see ref.~\cite{the3}) have been obtained for a kinetic model study 
of $J/\psi$ production at the LHC.

\begin{figure}[hbt]
\centering\includegraphics[width=.57\textwidth]{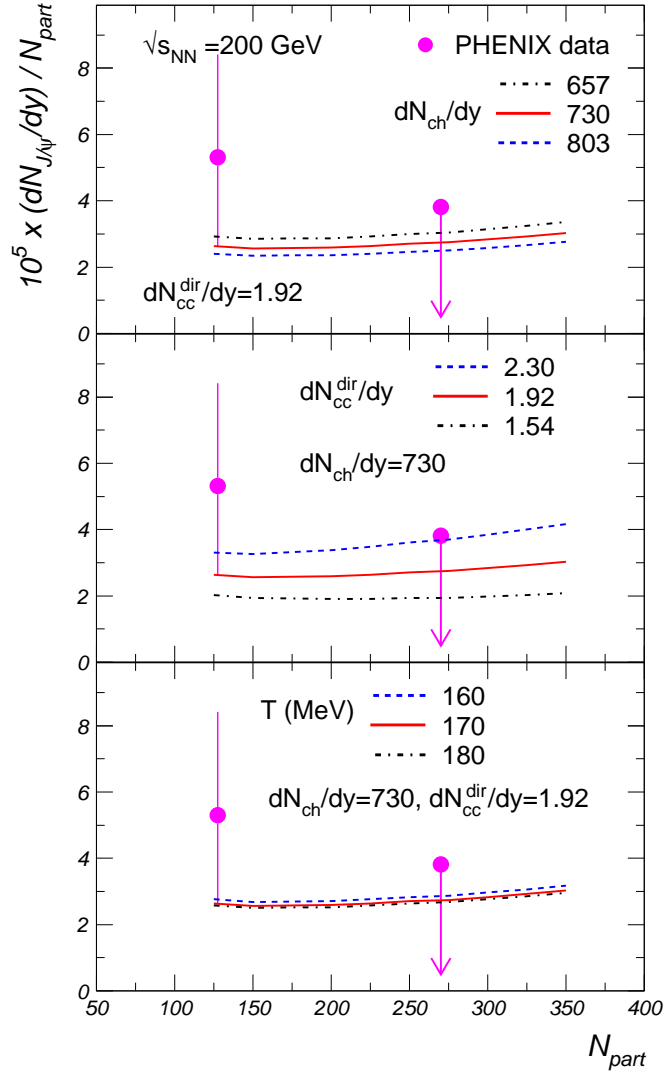}
\caption{Centrality dependence of rapidity densities of $J/\psi$
mesons at RHIC. Upper panel: sensitivity to $\ud N_{ch}/\ud y$; 
middle panel: sensitivity to $\ud N_{c\bar{c}}^{dir}/\ud y$; 
lower panel: sensitivity to $T$.  
The calculations are represented by lines.
The dots are experimental data from the PHENIX collaboration \cite{phenix}.
Note that the point for the central collisions is the upper limit
extracted by PHENIX for 90\% C.L. \cite{phenix}.
}
%with their statistical errors \cite{fra}.}
\label{aa:fig3}
\end{figure}

In Fig.~\ref{aa:fig3} we present the predicted centrality
dependence of the $J/\psi$ rapidity density normalized to
$N_{part}$ for RHIC energy ($\sqrt{s_{NN}}$=200 GeV).  
The three panels show its sensitivity on $\ud N_{ch}/\ud y$, 
$\ud N_{c\bar{c}}^{dir}/\ud y$, and (freeze-out) temperature $T$.  
The calculations are compared to experimental results of
the PHENIX Collaboration \cite{phenix}.
Note that the point for the central collisions is the upper limit
extracted by PHENIX for 90\% C.L. \cite{phenix}.
 The experimental data have been rescaled according to our
procedure \cite{dar} to calculate $N_{part}$ and the number of
binary collisions, $N_{coll}$.  Using the centrality intervals of
PHENIX, we obtain $N_{coll}$ = 275 and 782 (the differences to the
PHENIX values of 296 and 779, respectively, are small).  
We use $\sigma_{NN}$=42~mb and a Woods-Saxon density profile.

Within the still  large experimental error bars, the measurements
agree with our model predictions.  In Fig.~\ref{aa:fig3} only the
statistical errors of the mid-central data point are plotted. 
The systematic errors are also large \cite{phenix}.
%, similar to the statistical
%error for the higher centrality and are about half that for more
peripheral collisions. 
A stringent test of the present model can
only be made when high statistics $J/\psi$ data are available. 
In any case, very large $J/\psi$ suppression factors as predicted,
e.g., by \cite{vogt99} seem  to be not supported by the data.

We turn now to a more detailed discussion of the sensitivity of
our calculations to the various input parameters as quantified  in
Fig.~\ref{aa:fig3}.  First we consider  the influence of a $10\%$
variation of $\ud N_{ch}/\ud y$ on the centrality dependence of
$J/\psi$ yield.  Note that the total experimental uncertainty of
$\ud N_{ch}/\ud\eta$ (which is for the moment the measured
observable for most experiments) at RHIC is below 10\%
\cite{pho,bra}.
The sensitivity on the $\ud N_{ch}/\ud y$ values stems from the volume
into which the (fixed) initial number of charm quarks is
distributed. The smaller the particle multiplicities and thus also
the fireball volume, the more probable it is for charm quarks and
antiquarks to combine and form quarkonia. That is why one sees, in
the top panel in Fig.~\ref{aa:fig3}, that the $J/\psi$ yield is
increasing with decreasing charge particle multiplicity.

The sensitivity of the predicted $J/\psi$ yields on 
$\ud N_{c\bar{c}}^{dir}/\ud y$ is  also straightforward. The larger this 
number is in a fixed volume the larger is the yield of charmed hadrons. 
In case of charmonia the dependence on $\ud N_{c\bar{c}}^{dir}/\ud y$ 
is non-linear due to their double charm quark content, as reflected 
by the factor $g_c^2$ in equation (\ref{aa:eq2}). 
To illustrate the sensitivity of the model predictions on 
$\ud N_{c\bar{c}}^{dir}/\ud y$, we exhibit the results of a 20\% variation
with  respect to the value  given in Table~\ref{aa:tab1}. The open
charm cross section is not yet measured at RHIC. However, some
indirect measurements can be well reproduced, within the
experimental errors, by PYTHIA calculations using a p--p charm
cross sections scaled with the number of collisions $N_{coll}$ of
650 $\mu$b \cite{aver}. The corresponding value at
$\sqrt{s_{NN}}$=130 GeV is 330 $\mu$b \cite{phe}. For comparison,
the NLO pQCD values we are using \cite{vogt1} are 390 and 235 $\mu$b, 
respectively. Despite the still large experimental
uncertainties, this discrepancy, recognized earlier in ref.
\cite{the2}, needs to be understood.  We note that, dependent on
the input parameters used in the  NLO calculations
\cite{vogt1,the2}, possible variations of the open charm production 
cross section for the RHIC energy are of the order of $\pm$20\%. 
In terms of our model this variation corresponds 
to about a $\pm$30\% change  in $J/\psi$ yield  which is also
centrality dependent (see middle panel in Fig.~\ref{aa:fig3}). 
If we use the PHENIX p--p cross section of 650 $\mu$b, the
calculated yield is a factor 2.5 larger for $N_{part}$=350 and
increases somewhat stronger with centrality, in disagreement
with the data.  As apparent in Fig.~\ref{aa:fig3}, the
predictive power of our model or of any similar model relies
heavily on the accurate knowledge of the overall charm production
cross section.  A simultaneous description of the centrality
dependence of open charm together with $J/\psi$ production is, 
in this respect, mandatory to test the 
concept of the statistical origin of open and hidden charm 
hadrons in heavy ion collisions at relativistic energies.
%predictions of the model.

\begin{table}[htb]
\caption{Temperature dependence at RHIC, for central collisions.} 
\label{aa:tab3}
\begin{tabular}{c|cccc}
$T$ & 160   & 170 & 180 & 180/160 \\ \hline
$\ud N_{D^+}^{th}/\ud y$ & 0.0368 &0.0554 & 0.0775 & 2.106\\
$\ud N_{J/\psi}^{th}/\ud y$ &
0.672$\cdot$10$^{-4}$ & 0.153$\cdot$10$^{-3}$ & 0.311$\cdot$10$^{-3}$
& 4.628 \\
$g_c$ & 12.85 & 8.33 & 5.77 & 1/2.227 \\
$\ud N_{D^+}/\ud y$ & 0.415 & 0.404 & 0.391 & 0.942 \\
$\ud N_{J/\psi}/\ud y$ & 0.0111 & 0.0106 & 0.0103 & 0.928 \\
$N_{\psi'}/N_{J/\psi}$ & 0.031 & 0.037& 0.045 & 1.452 \\ 
\end{tabular}
\end{table}

The apparent weak dependence of $J/\psi$ yield on  freeze-out
temperature, seen in  Fig.~\ref{aa:fig3}, may be surprising. This
is particularly the case since in the thermal model considered in
\cite{marek} this dependence was shown to be very strong.  In our
model this result is a consequence of  the charm balance equation
(\ref{aa:eq2}).  The temperature variation leads, obviously, to a
different number of thermally produced charmed hadrons as in
\cite{marek}, but this is compensated by the $g_c$ factor.
The approximate temperature dependence of $g_c$ and the $J/\psi$ yield are:
\begin{equation}
g_c(T) \sim 1/N_D^{th} \sim e^{\frac{m_D}{T}}, \quad
N_{J/\psi}(T)=g_c^2 N_{J/\psi}^{th} \sim e^{\frac{2m_D-m_{J/\psi}}{T}}
\end{equation}
As a result of the small mass difference in the exponent  the
$J/\psi$ yield exhibits  only a weak sensitivity on  $T$. This
is in contrast to the purely thermal case where the yield  scales
with $\exp(-m_{J/\psi}/T)$.
The results are summarized in  Table~\ref{aa:tab3}, where  we
present  the number of $J/\psi$ and 
$D^+$ mesons produced thermally ($N_{J/\psi}^{th}$ and
$N_{D^+}^{th}$, respectively) and  within the frame of our  statistical
hadronization model. Also shown are the values of $g_c$ and the yield 
ratio $\psi'/J/\psi$ for three different values of $T$. 
The compensations are evident in the last column of Table~\ref{aa:tab3}, 
where ratios for the two extreme temperatures are calculated. 
The only exception is the ratio $\psi'/J/\psi$, which is obviously 
identical in the statistical hadronization scenario and in the thermal 
model and coincides, for T$\simeq$170 MeV, with the measured value at 
SPS \cite{pbm1}.

\begin{figure}[hbt]
%exec plot#jpsi-npart-rhic3
\centering\includegraphics[width=.59\textwidth]{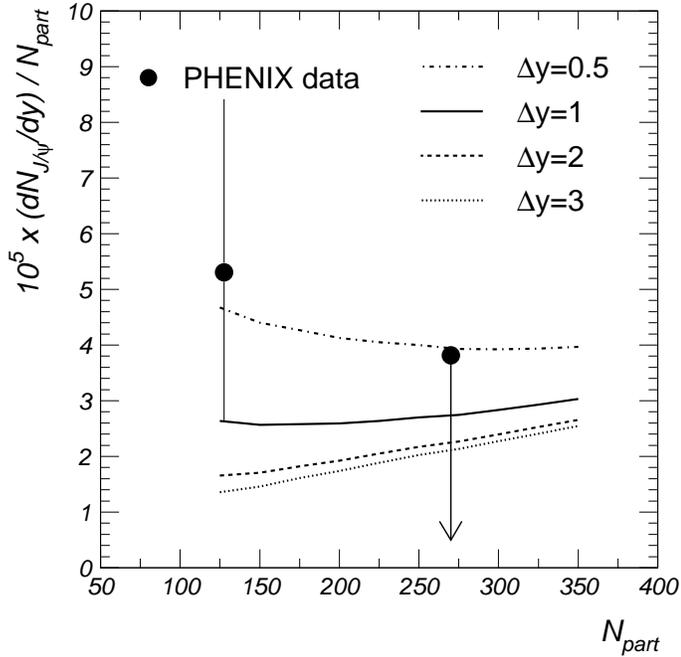}
\caption{Centrality dependence of rapidity densities of $J/\psi$
mesons at RHIC for different rapidity window sizes. 
The lines are calculations, the dots are experimental data from 
PHENIX collaboration \cite{phenix} (the point for the central collisions 
is the upper limit for 90\% C.L.).
}
\label{aa:fig4}
\end{figure}

Most of our results presented above are obtained considering a one
unit rapidity window at midrapidity, while results for the full volume
were presented only for the SPS. Unlike the kinetic model of
Thews et al. \cite{the1,the2}, our model does not contain
dynamical aspects of the coalescence process.  However, in our
approach, the width of the rapidity window does influence the
results in the canonical regime. For the grand-canonical case,
attained only at LHC energy, there is no dependence on the width
of the rapidity window, due to a simple cancellation between the
variation of the volume, proportional to the rapidity slice in
case of a flat rapidity distribution, and the variation of
$N_{c\bar{c}}^{dir}$, also proportional to the width of the
rapidity slice.  In Fig.~\ref{aa:fig4} we present the centrality
dependence of $J/\psi$  rapidity densities for RHIC energy  and
for different rapidity windows $\Delta$y from 0.5 to 3. The
dependence on $\Delta$y resembles that of the kinetic model
\cite{the2}, but is less pronounced. The available  data  are not
yet precise enough to rule out any of the scenarios considered.
However, for the kinetic model \cite{the2}, the cases of small $\Delta$y 
seem to be ruled out by the present PHENIX data.

We stress in this context that the size of $\Delta$y window has a
potentially large impact on the  results at  SPS energy. It is
conceivable that no charm enhancement is needed to explain the
data if one considers a sufficiently narrow rapidity window
for the statistical hadronization.

The  results presented  above were obtained under the assumption
of statistical hadronization of quarks and gluons.
In addition, for charm quarks the yields  were constrained by the charm 
balance equation that was formulated in the canonical ensemble. 
We have further assumed that charm quarks are entirely produced via 
primary hard scattering  and thermalized in the QGP. No secondary 
production of charm in the initial and final state was included in our 
calculations.
Final state effects like nuclear absorption of $J/\psi$ \cite{fnal} are 
also neglected.

\section{Conclusions}
We have demonstrated that the statistical coalescence approach yields
a good description of the measured centrality dependence of $J/\psi$
production at SPS energy, albeit with a charm cross section increased
by a factor of 2.8 compared to current NLO perturbative QCD
calculations. Rapidity densities for open and hidden charm mesons are
predicted to increase strongly with energy, with striking changes in
centrality dependence. First RHIC data on $J/\psi$ production support
the current predictions, although the experimental errors are for the 
moment too large to allow firm conclusions.  
We emphasize that the predictive power of our model relies heavily on 
the accuracy of the charm cross section, not yet measured in heavy-ion 
collisions.  The statistical coalescence implies travel of charm quarks 
over significant distances e.g. in a QGP. If the model predictions will 
describe consistently precision data this would be a clear signal for 
the presence of a deconfined phase.

\section*{Acknowledgments}
We acknowledge stimulating discussions with  R.L. Thews.
K.R. acknowledges the support of the Alexander von Humboldt Foundation (AvH).

\end{document}